\documentclass[preprint,12pt,authoryear]{elsarticle}

\usepackage{amssymb}
\usepackage{amsmath}
\usepackage{amsthm}
\usepackage{multirow}
\usepackage{xcolor}
\usepackage[most]{tcolorbox}
\usepackage{todonotes}
\usepackage{placeins}
\usepackage{float}
\usepackage{url}
\newtcolorbox{graybox}{
  colback=gray!20,   
  colframe=gray!20, 
  boxrule=0pt,      %
  arc=0pt,          %
  left=8pt,
  right=8pt,
  top=6pt,
  bottom=6pt,
  boxsep=0pt,
  breakable
}

\usepackage{listings}
\lstdefinestyle{promptstyle}{
    basicstyle=\small\ttfamily\color{black},
    breaklines=true,
    frame=single
}

\begin{document}

\begin{frontmatter}



\title{Could Large Language Models work as Post-hoc Explainability Tools in Credit Risk Models?} 


\author[author1]{Wenxi Geng} 
\author[author2]{Dingyuan Liu}
\author[author3]{Liya Li}
\author[author4]{Yiqing Wang\corref{cor1}} 

\cortext[cor1]{Corresponding Author}
\ead{woshilucy712@gmail.com}
\affiliation[author1]{organization={Citigroup},
            city={Irving},
            state={TX},
            country={United States}}
\affiliation[author2]{organization={University of North Carolina at Chapel Hill},
            city={Chapel Hill},
            state={NC},
            country={United States}}
\affiliation[author3]{organization={Georgia Institute of Technology},
            city={Atlanta},
            state={GA},
            country={United States}}
\affiliation[author4]{organization={Southern Methodist University},
            city={Dallas},
            state={TX},
            country={United States}}

\begin{abstract}

Large language models (LLMs) have shown promise in translating model-based explanations into human-readable narratives. This study evaluates whether LLMs can serve as post-hoc explainability interfaces for credit risk models, focusing on their ability to preserve feature-importance rankings and generate autonomous explanations. Using a LendingClub dataset, we compare LLM outputs with SHAP and coefficient-based attributions on three major LLMs, including  \textit{GPT-4-turbo}, \textit{Claude-Sonnet-4.5}, and \textit{Gemini-2.5-Flash}. Results indicate that LLMs reliably reproduce reference rankings under controlled prompts but show limited alignment when generating explanations autonomously. These findings suggest that LLMs are best deployed as narrative interfaces rather than substitutes for formal attribution methods in credit risk governance.

\end{abstract}





\end{frontmatter}


\section{Introduction}
\label{SecIntro}

Rapid technological advances have enabled FinTech lenders to become major participants in the U.S. personal-loan market alongside traditional institutions. By 2022, FinTech-owned loans accounted for approximately 14\% of personal loans in the United States \citep{flagg2023fintech}. This rapid growth has been driven in large part by the adoption of advanced machine learning (ML) models that streamline underwriting processes, improve predictive performance, and enhance predictive accuracy. Despite these advantages, the increasing reliance on complex and opaque ML models raises significant regulatory and risk management concerns. Regulatory frameworks such as the Equal Credit Opportunity Act (ECOA) \citep{ECOA} and the Fair Credit Reporting Act (FCRA) \citep{FCRA} establish a series of requirements related to adverse action notices and the communication of reasons for certain credit decisions. Beyond regulatory compliance, explainability plays a critical role in fostering customer trust \citep{CustomerExperience}, allowing internal auditing \citep{Audit}, and supporting risk management practices \citep{aziz2019machine}.

To mitigate the opacity of ML models, financial institutions commonly adopt post-hoc explainable AI (XAI) methods such as SHapley Additive exPlanations (SHAP) and Local Interpretable Model-agnostic Explanations (LIME)\allowbreak\ \citep{lundberg2017shap,LIME}. These techniques quantify feature-level contributions by decomposing model predictions into additive attribution components. While effective at identifying important features, such methods primarily provide numerical outputs-such as feature importance ranking or contribution values which are difficult to interpret and operationalize in decision-making process. Furthermore, these mathematical outputs lack the flexibility to incorporate domain-specific knowledge or generate contextual narratives that align with how credit analysts reason about lending decisions. 

Recent advances in large language models (LLMs) offer a potential pathway to bridge this interpretability gap. LLMs have demonstrated remarkable capabilities in natural language understanding, reasoning, and generation in high-stake domains. Unlike traditional XAI approaches, LLMs may translate feature-level attributions into coherent, domain-informed explanations expressed in natural language. In principle, this capability positions LLMs as promising candidates for serving as narrative interfaces that render ML-based credit decisions more transparent and accessible to diverse stakeholders. However, whether LLMs can reliably perform such translation tasks—and whether they can autonomously generate explanations that align with model-grounded attributions—remains an open empirical question.

In this paper, we investigate the potential of LLMs as post-hoc explainers for credit risk predictions through in-context learning (ICL), focusing on the following research questions:
\begin{itemize}
    \item \textbf{RQ1 (LLM as Translator)}: To what extent can large language models accurately preserve feature-importance rankings when producing structured ranked feature lists, given reference attributions produced by established XAI methods (e.g., SHAP or logistic regression coefficients)? Specifically, can LLMs maintain the integrity of the ranked structure embedded in model-grounded reference explanations under strict output constraints?

    \item \textbf{RQ2 (LLM as Autonomous Explainer)}: To what extent can large language models autonomously infer feature-importance rankings that align with model-grounded reference explanations (e.g., SHAP or logistic regression), when such rankings are not provided as inputs and how few-shot prompting influences this alignment?
\end{itemize}

We find that, in a controlled setting where reference rankings are provided, LLMs almost always preserve the correct Top-$K$ feature set, with only rare deviations. In contrast, when reference rankings are not supplied, alignment with model-grounded reference explanations is generally low. Few-shot prompting improves alignment for the logistic regression but does not consistently enhance performance for XGBoost, suggesting that LLMs struggle to recover nonlinear, interaction-driven reasoning from prompt cues alone. Overall, these results indicate that LLMs are best positioned as narrative interfaces grounded in auditable attribution rankings, rather than as substitutes for post-hoc explainers in credit lending model risk management. This distinction underscores the importance of maintaining formal attribution methods as the primary source of explanatory accountability.

The remainder of the paper is organized as follows. Section 2 reviews the related literature on explainable AI and the use of LLMs in financial applications. Section 3 describes the experimental design, including data processing, baseline model training, prompt engineering strategies, and evaluation metrics. Section 4 presents the empirical results for both research questions. Section 5 concludes with a discussion of implications and directions for future research.


\section{Literature Review}
\label{SecReview}

Credit scoring has historically relied on transparent models such as logistic regression and rule-based decision trees due to their interpretability, ease of validation, and compatibility with regulatory requirements \citep{thomas2000survey}. In recent years, machine learning models—including support vector machines (SVMs) and neural networks (NNs)—have been increasingly applied to credit risk modeling because of their ability to capture nonlinear relationships and achieve superior predictive performance \citep{baesens2003benchmarking, lessmann2015benchmarking}. Despite their predictive advantages, the complexity and black-box nature of these models introduce substantial challenges for interpretability, validation, and regulatory compliance.

To address these challenges, post-hoc explainable AI (XAI) methods such as SHAP and LIME have been widely adopted to interpret black-box credit models and to support feature-based explanation workflows in credit risk management. In regulated credit settings, such outputs may inform adverse-action review processes, but they do not by themselves establish legal sufficiency without additional governance, validation, and compliance controls \citep{misheva2021explainable}. By estimating marginal feature contributions, those methods provide a practical mechanism for attributing predictions to input variables. However, their results are primarily numerical—such as feature importance scores or contribution values and may lack semantic clarity. As a result, they can be difficult for non-technical stakeholders to interpret and may not fully support human-centered decision processes \citep{tambwekar2023towards}.

Recent studies further show the growing use of LLMs in high-stakes decision-support settings, including clinical decision support, education, and legal applications \citep{cascella2023evaluating, han2024llm, siino2025exploring}. In critical care medicine, Shi et al. \citep{shi2025large} reviewed recent LLM applications and highlighted both their potential for clinical decision support and persistent risks such as hallucination, prompt sensitivity, bias, and privacy concerns. Building on this broader trend, recent work has also explored tool-augmented and multi-agent LLM frameworks to improve reliability in clinical tasks \citep{goodell2025large, shi2025aki}. While these studies suggest that LLMs may serve as useful explanatory or decision-support interfaces, they also highlight persistent concerns related to hallucination, prompt sensitivity, interpretability, bias, and privacy. These concerns motivate the need to evaluate whether LLM-generated explanations remain faithful and reliable in high-stakes domains such as credit risk management.

Kroeger et al. \citep{kroeger2023large} examined the feasibility of using LLMs as post-hoc explainers by evaluating their ability to reproduce feature-importance rankings. However, their experimental design relied on synthetic feature representations lacking domain-specific semantic content, thereby primarily assessing symbolic ranking reproduction rather than explanations grounded in real-world decision contexts. Mohanty et al. \citep{mohanty2025impact} further demonstrated that LLMs, when guided by few-shot prompting, can reproduce attribution patterns similar to classical post-hoc methods in a multi-disease prevention setting.

Despite these emerging findings, the literature does not yet provide a systematic understanding of how LLM-based explanations behave in the credit risk domain. In particular, it remains unclear whether LLMs can reliably preserve structured feature-importance rankings when model-grounded reference explanations are explicitly provided, or whether they can autonomously infer such structures in the absence of reference attributions. These two settings correspond to distinct deployment scenarios for LLM-based explanation systems in credit risk management and directly motivate our investigation of LLMs as translators versus autonomous explainers.

\section{Methods}
\label{SecMethods}
\begin{figure}
	\centering
	\includegraphics[width=\textwidth]{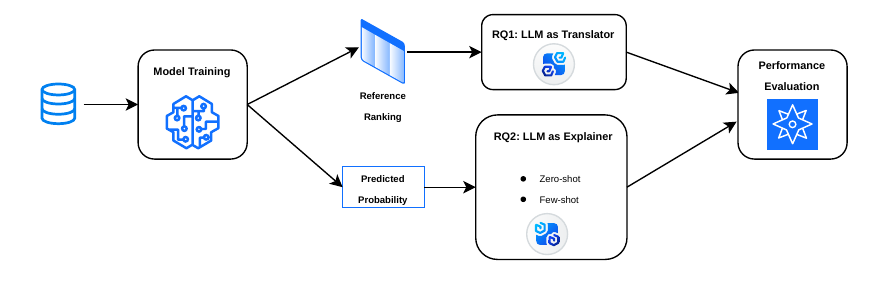}
	\caption{Overall framework of the proposed LLM-based attribution study}
	\label{fig:method_flowchart}
\end{figure}

Figure \ref{fig:method_flowchart} summarizes the overall framework of this study. We first train two baseline credit risk models—logistic regression and XGBoost—and derive model-grounded reference attributions (logistic regression coefficient-based contributions and SHAP values for XGBoost), which are converted into feature-importance rankings. These reference rankings and/or model outputs are then provided to LLMs to address RQ1 and RQ2. 

For RQ1, LLMs are evaluated as translators conditioned on the reference ranking. For RQ2, we evaluate both zero-shot and few-shot prompting configurations. In both settings, LLM-generated feature rankings are subsequently compared with the reference rankings to quantify alignment and consistency. The following subsections describe each step in detail.

\subsection{Data Structure and Cleaning}
\label{SubData}

This study utilizes a publicly available LendingClub  \citep{lendingclub_2007_11} consumer loan dataset obtained from the data.world repository\footnote{The dataset was downloaded from the data.world repository: \url{https://data.world/jaypeedevlin/lending-club-loan-data-2007-11}, which contains anonymized information on borrower characteristics, loan contract terms, credit bureau attributes, and loan performance outcomes for loans originated between 2007 and 2011.}

To ensure suitability for predictive modeling and explainability analysis, a systematic data pre-processing pipeline was applied. First, variables with all missing values or zero variance are excluded, as they provide no discriminatory power for modeling. Second, categorical features exhibiting sparse or excessively granular levels are consolidated into economically meaningful groups based on their underlying semantic interpretation, thereby reducing dimensionality while preserving information content. Third, features that were either unrelated to loan performance or recorded after loan origination are dropped to prevent data leakage and ensure temporal validity of the predictive model.

Categorical variables were also encoded according to their measurement scales. Ordinal categorical variables, such as credit grade, are mapped to integer encoding that preserves their inherent ranking structure and economic ordering. Nominal categorical variables are transformed through one-hot encoding to avoid imposing artificial ordinal relationships among category levels. These encoding strategies support both predictive performance and the interpretability of subsequent post-hoc explanations.

Following these pre-processing procedures, the final modeling dataset comprises 39,239 loan applicants with 24 independent features. The target variable is a binary indicator of loan outcome, where loans classified as Charged Off are labeled as 1 and loans classified as Fully Paid are labeled as 0. The non-default-to-default ratio is approximately 6.15:1, indicating substantial class imbalance. 

For notational clarity in subsequent sections, we formalize the problem setup as follows: Let $\{(x_i,y_i)\}_{i=1}^{N}$ denote the complete loan dataset, where $N=39,239$, input vector $x_i\in \mathbb{R}^{m}$ with $m = 24$ represents the independent feature vector for loan $i$, and $y_i \in \{0,1\}$ is the corresponding binary outcome. A predictive model $f(\cdot)$ maps the independent feature vector of each loan to a probability of default: $p_i=f(x_i)$. 

\subsection{Baseline Prediction Model}
\label{SubModel}
The predictive task is formulated as a binary classification problem, in which the model $f(\cdot)$ estimates the probability that a loan will default. To provide a prediction backbone for subsequent post-hoc explainability analysis, we train two representative credit risk models: logistic regression and eXtreme Gradient Boosting (XGBoost) \citep{Xgboost}. 

Logistic regression serves as an inherently interpretable baseline model, offering coefficient-based attributions that directly reflect linear feature contributions. In contrast, XGBoost represents a state-of-the-art tree-based ensemble method capable of capturing nonlinear relationships and complex feature interactions, and is widely adopted in industry practice. The inclusion of both models enables a comparative evaluation of LLM behavior across linear and nonlinear predictive structures.

Following data pre-processing and variable selection, the logistic regression model was calibrated by tuning the L2 regularization hyperparameter $\lambda$. The area under the Precision-Recall Curve (PR-AUC) \citep{prauc} was adopted as the optimization criterion to better capture model performance on the minority class under severe class imbalance inherent in the dataset. For the XGBoost model, hyperparameter tuning was performed using stratified five-fold cross-validation, with PR-AUC serving as the evaluation metric. This evaluation choice ensures consistent performance assessment across both linear and nonlinear modeling frameworks.

All models were trained on the preprocessed dataset described in Section \ref{SubData}. The data were randomly partitioned into training and test subsets using a $70:30$ split ratio, stratified by the target variable to maintain class distribution. Model fitting was performed exclusively on the training subset, while all subsequent explainability analyses were conducted on the held-out test set to ensure that explanations reflect model behavior on previously unseen observations and avoid overfitting-induced bias in attribution patterns.

Model performance was evaluated using three complementary metrics, each capturing distinct aspects of predictive quality under imbalanced classification settings. 

\begin{itemize}
    \item \textbf{PR-AUC}: The area under the Precision–Recall curve evaluates the model's ability to correctly rank minority-class observations by capturing the precision–recall trade-off across all decision thresholds. Unlike the area under the Receiver Operating Characteristic curve (AUROC), PR-AUC is more informative under class imbalance, as it explicitly emphasizes performance on the positive (default) class. Accordingly, PR-AUC is adopted as the primary metric for assessing discriminatory performance in this study.

    \item \textbf{Macro-averaged F1-score} \citep{F1score}: Computed as the unweighted average of class-specific F1-scores, this metric assigns equal importance to both majority and minority classes. It provides a balanced evaluation of classification performance and mitigates the dominance of the majority class in aggregate accuracy measures.

    \item \textbf{Kolmogorov–Smirnov (KS) Statistic} \citep{KS}: The KS statistic measures the maximum vertical distance between the cumulative distribution functions of predicted risk scores for the positive and negative classes. Widely used in credit risk modeling, it quantifies the model’s discriminatory power. Higher KS values indicate stronger separation between classes and thus superior model discrimination.
     
\end{itemize}

\subsection{Reference Explainability Methods}
\label{SubSHAP}
Given the distinct characteristics of logistic regression and XGBoost, we adopt model-specific explanation approaches to construct model-grounded reference explanations for comparison with LLM-generated outputs. 

For the logistic regression model specified in Equation \ref{eq:logit}, the regression coefficients $\beta$ provide a natural and interpretable measure of feature influence under a linear log-odds specification. Instance-specific feature contribution is quantified by the product $\beta_j x_{ij}$, which combines the estimated coefficient ($\beta_j$) with the corresponding feature value ($x_{ij}$) for each observation $i$. The sign of this product indicates the directional effect of the feature on predicted default risk, where a positive value increases the log-odds of default and a negative value decreases it. The magnitude reflects the strength of the contribution to the linear predictor. The overall prediction is obtained by summing these contributions across all $m$ features and adding the intercept term. This additive decomposition provides a complete and exact attribution of the model's prediction to individual features. 

\begin{equation}
\label{eq:logit}
\log\!\left( \frac{p_i}{1-p_i} \right)
= \beta_0 + \sum_{j=1}^{m} \beta_j x_{ij}
\end{equation}

For the XGBoost model, we employ SHAP to decompose model output into feature-level contributions for each instance. SHAP is a unified framework grounded in cooperative game theory that assigns each feature an importance value given a particular prediction. While the details of the SHAP methodology are beyond the scope of this paper, we provide the key formulation below for completeness.

For each instance $i$, we define the SHAP attribution vector $\boldsymbol{\phi}_i = (\phi_{i1},\ldots,\phi_{im})$, where $\phi_{ij}$ represents the signed contribution of feature $j$ to the model output for observation $i$, relative to a baseline $\phi_0$. SHAP satisfies an additive decomposition of the model output,  \begin{equation}
\label{eq:shap_additive}
f(\mathbf{x}_i) = \phi_0 + \sum_{j=1}^{m} \phi_{ij},
\end{equation}
where \( f(\mathbf{x}_i) \) denotes the model output. Similar to the logistic regression, the sign of $\phi_{ij}$ indicates whether feature $j$ increases or decreases the predicted default risk relative to the baseline expectation, while its magnitude quantifies the strength of that contribution. 

Both  $\beta_j x_{ij}$ and $\phi_{ij} $  are computed in the encoded feature space used by the predictive models. To ensure interpretability at the original feature level, we subsequently aggregate attributions for each categorical feature by grouping its one-hot components. For a categorical feature \( g \) with one-hot encoded components indexed by \( \mathcal{J}(g) \), the grouped local contribution is defined as $\sum_{j \in \mathcal{J}(g)} \phi_{ij} \text{ or } \sum_{j \in \mathcal{J}(g)} \beta_j x_{ij}$, dependening on the model algorithm. This aggregation ensures that feature importance rankings are evaluated at the semantically meaningful variable level rather than at the level of encoded dummy variables.

For example, the original feature \textit{Home Ownership} is expanded by one-hot encoding into four transformed columns: \textit{Mortgage}, \textit{Own}, \textit{Rent}, and \textit{Others}. These four columns are then grouped back to the original feature \textit{Home Ownership} in both attribution pipelines. Logistic regression aggregates the corresponding transformed-level \(\beta x\) terms, while XGBoost aggregates the corresponding transformed-level SHAP values using the same group indices.


\subsection{LLM-based Post-hoc Explainability}
\label{SubLLM}
In-context learning (ICL) \citep{brown2020language} refers to the ability of large language models to perform tasks by conditioning on task instructions or demonstration examples provided directly within the prompt, without requiring parameter updates or additional fine-tuning. This capability enables LLMs to generate explanations under different prompting strategies, including zero-shot settings with task instructions only and few-shot settings augmented with demonstration examples. 

To address the two research questions and to evaluate LLM performance under different prompting configurations, we employ three widely used commercial LLMs: \textit{GPT-4-turbo}, \textit{Claude-Sonnet-4.5}, and \textit{Gemini-2.5-Flash}. All models are accessed via the \textit{LiteLLM} framework (version 1.81.3), which provides a unified interface for model invocation and experiment management. To enhance reproducibility and reduce stochastic variation across runs, the temperature parameter is fixed at 0 for all models. This setting allows controlled sampling while maintaining consistency across experimental conditions.

All experiments are conducted in Python (version 3.12.12) using standard scientific computing libraries, including pandas (2.2.2), NumPy (2.0.2), scikit-learn (1.6.1), SHAP (0.50.0), XGBoost (3.1.3), and Optuna (4.6.0), within the Google Colab environment. The complete prompt templates used in each experiment are provided in Appendix~\ref{sec:prompt}.

\subsubsection{RQ1: LLM as Translator}
\label{Methods_RQ1}

RQ1 evaluates whether LLMs can preserve model-grounded attribution rankings when reference attributions are explicitly provided within the prompt. Importantly, RQ1 does not assess the model’s ability to independently infer feature importance from raw input variables. Rather, it isolates a more fundamental capability that is essential for credit risk applications: the ability of an LLM to faithfully preserve and reproduce a specified attribution structure under conditions of minimal ambiguity.
This controlled setting enables a direct assessment of structural fidelity between the reference attributions and the LLM-generated explanations. The experimental procedure is formalized as follows.

For each instance $(x_i,y_i)$ and a trained baseline predictive model $f(\cdot)$, we implement the following procedure:

\begin{itemize}

    \item \textit{Step 1: Instance-Level Information Preparation}:  
    For each instance \( i \), we construct a structured input consisting of the observed outcome \( y_i \), the predicted probability \( p_i = f(\mathbf{x}_i) \), and an attribution table containing the reference ranked feature set $\mathcal{T}_i$ derived from the baseline model's explanation mechanism. The set $\mathcal{T}_i$ represents the model-grounded attribution structure that the LLM is expected to reproduce.
    
    \item  \textit{Step 2 Prompt Construction with Strict Output Constraints}: We design a structured prompting protocol that explicitly constrains the LLM's output format. In particular, the prompt instructs the LLM to return a ranked list of features drawn from the original input space. No additional inferred features or external knowledge are permitted, ensuring that the task is framed as controlled reproduction rather than independent feature inference. 
    
    \item \textit{Step 3 LLM Invocation and Explanation Generation}: Using the constructed prompt, we query the LLM to generate an instance-specific ranked feature list $\hat{\mathcal{T}}_i$ for each instance. This procedure is repeated for each of the three LLMs under identical experimental conditions to enable consistent comparison of ranking-preservation fidelity across models. 
    
    \item \textit{Step 4 Instance-level Alignment Evaluation}: The LLM-generated ranking $\hat{\mathcal{T}}_i$ is compared against the reference ranking set $\mathcal{T}_i$ at multiple top-$K$ cutoffs. Ranking-based alignment metrics, described below, are used to quantify how accurately the LLM preserves both the feature set and relative ordering provided in the prompt.

\end{itemize}

This algorithm evaluates whether an LLM can faithfully reproduce a reference explanation when the attribution structure is explicitly supplied and sources of ambiguity are minimized. For each test instance, the LLM was instructed to return an explicit ranked list of features, and the returned ranking was directly used for computing the ranking-based evaluation metrics without additional post-processing, synonym mapping, duplicate correction, missing-feature imputation, or tie-resolution procedures.

Before evaluating LLM performance, we first assess schema-level fidelity by checking whether each returned feature belongs to the supplied feature list, whether duplicate features appear, whether the required number of ranked features is returned, and whether the response follows the required JSON structure. This check ensures that the output satisfies the constrained format before Top-K overlap and Kendall’s $\tau$ are used to evaluate ranking preservation.

Because the RQ1 output schema requires only a ranked list of feature names in JSON format, the generated outputs do not include free-form narratives, directionality statements, or causal claims. Therefore, narrative-faithfulness checks, such as unsupported causal language or incorrect directional interpretation, are not directly applicable to the current RQ1 design.

To quantify LLM performance under this controlled setting, we consider two complementary metrics: 
\begin{itemize}
    \item Top-$K$ Feature Overlap: This metric measures the proportion of features shared between the LLM-generated and reference top-$K$ feature sets, as formalized in formula \ref{eq:top_k_overlap}. It captures whether the same top-K features are identified, regardless of their relative ordering. 
    \begin{equation}
    \label{eq:top_k_overlap}
    \mathrm{Overlap@K}_i
    =
    \frac{\left| \hat{T}_i^{(K)} \cap T_i^{(K)} \right|}{K}
    \end{equation}
    where $T_i^{(K)}$ denotes the top-$K$ features from the reference and $\hat{T}_i^{(K)}$ is the corresponding LLM-generated set for instance $i$. 

    \item Kendall's $\tau$ coefficient \citep{kendallchat_stats}: measures the degree of pairwise concordance between the reference ranking  $\mathcal{T}_i$ and the LLM-generated ranking $\hat{\mathcal{T}}_i$. It is computed over the intersection of the reference Top-K and LLM-generated Top-K feature sets, and therefore evaluates whether the LLM preserves the relative ordering among features that appear in both Top-K lists.
    
\end{itemize}

\subsubsection{RQ2 LLM as Autonomous Explainer}
\label{Methods_RQ2}

RQ2 examines whether LLMs can autonomously generate meaningful feature-importance rankings in the absence of explicit reference guidance. This capability is particularly relevant for practical deployment scenarios in which LLMs function as standalone explanation tools or where model-based attribution rankings are not directly accessible.

In addition, RQ2 evaluates the impact of in-context learning (ICL) strategies by comparing the quality of LLM-generated explanations under zero-shot and few-shot prompting configurations. This analysis assesses whether demonstration-based prompting improves alignment with model-grounded reference explanations when structural guidance is not explicitly provided.

To further distinguish practical explanation performance from full attribution-structure recovery, we evaluate RQ2 under two output conditions. In the \textit{Top-4 condition}, the LLM is instructed to return only the four most important features. This condition reflects practical credit explanation and adverse-action-related review settings, where explanations typically focus on a small number of primary contributing factors rather than a complete ranking of all input variables. In the \textit{full-ranking condition}, the LLM is instructed to return a complete ranking of all input features. This condition provides a stricter evaluation of whether the LLM can recover the full model-grounded attribution structure.

For each instance $(x_i, y_i)$, each trained baseline predictive model $f(\cdot)$, and each output condition, we implement the following procedure:

\begin{itemize}

    \item \textit{Step 1: Instance-Level Information Preparation}: 
    For each instance \( i \), we construct a structured input consisting of the feature vector \( \mathbf{x}_i \), the observed outcome \( y_i \), and the predicted probability \( p_i = f(\mathbf{x}_i) \). Unlike RQ1, no reference attribution ranking is provided at this stage.
    
    \item \textit{Step 2 Prompt Construction with ICL Strategy Variation}: Following the output format constraints established in RQ1, we construct prompts by varying two experimental dimensions: the in-context learning strategy and the required ranking length. For the in-context learning strategy, we consider two configurations:

    \begin{itemize}
        \item \textit{Zero-shot Setting:} The prompt contains only the instance information and task instructions, without any demonstration examples.
        \item \textit{Few-shot Setting:} Besides the instance information and task instructions, the prompt also includes two demonstration examples along with their reference feature rankings derived from the baseline model's explanation mechanism. We select two extreme cases in the training set, corresponding to observations with the highest and lowest predicted probabilities of default. This selection was intended to provide the LLM with clear contrasting examples of how application-level information should be translated into ranked feature explanations.
    \end{itemize}
    For each prompting configuration, we evaluate both the \textit{Top-4 condition} and the \textit{full-ranking condition} as described above.
    
    \item \textit{Step 3 LLM Invocation and Ranking Generation}: Using the constructed prompt, we query the LLM to generate a ranked feature $\hat{\mathcal{T}}_i$ for each instance. The process is repeated across all three LLMs under zero-shot and few-shot configurations. 

    \item \textit{Step 4 Instance-level Alignment Evaluation}: The LLM-generated ranking $\hat{\mathcal{T}}_i$ is compared against the reference ranking $\mathcal{T}_i$ at multiple top-$K$ cutoffs using the same metrics described in Section \ref{Methods_RQ1}
\end{itemize}

This procedure evaluates whether LLMs can independently generate feature-importance rankings that align with model-grounded reference explanations in the absence of explicit attribution guidance. By systematically comparing zero-shot and few-shot configurations, we assess the extent to which demonstration examples enhance the LLM’s capacity to identify relevant features and approximate the underlying importance structure of the predictive model.


\section{Results}
\label{SubResults}

\subsection{Baseline Model Performance}
To evaluate the predictive performance of the baseline models, we consider several key metrics, including PR-AUC, macro-averaged F1-score, and KS statistics as described in Section~\ref{SubModel}. Table~\ref{tab:baseline_performance} reports the validation sample performance of the logistic and XGBoost models. 

The logistic regression model achieves a PR-AUC of $0.27$ and a KS statistic of $28.9\%$, indicating limited discriminatory power in separating defaulted from paid-off accounts. In contrast, the XGBoost model demonstrates substantially stronger predictive performance, with a higher PR-AUC of 0.38 and a higher KS statistic of $40.4\%$, reflecting improved ranking and class separation capabilities.

\begin{table}[H]
\centering
\caption{Key Metrics of Logistic and XGBoost Classification}
\label{tab:baseline_performance}
\begin{tabular}{lccc}
\hline
Model & PR-AUC & Macro-averaged F1-score & KS \\
\hline
Logistic  & 0.27 & 0.46 & 28.79 \\
XGBoost & 0.38 & 0.63 & 40.4 \\
\hline
\end{tabular}
\end{table}

\subsection{RQ1: LLM as Translator}
We evaluate RQ1 to examine whether LLMs can faithfully preserve instance-level reference explanations $\mathcal{T}_i$ when the attribution ranking is explicitly provided and output ambiguity is minimized. The evaluation is conducted on a stratified subset of 500 test observations, with 125 instances randomly sampled from each prediction category: true positives, true negatives, false positives, and false negatives. This design allows the ranking-preservation task to be assessed across both correctly and incorrectly classified cases.

Before evaluating ranking-preservation metrics, we first performed a schema-level fidelity check. This check evaluates whether each RQ1 output conformed to the required output structure, including whether all returned feature names appeared in the supplied feature list, whether duplicate features were present, whether the required ranked positions were returned, and whether the response conformed to the required JSON format. As shown in Table~\ref{tab:rq1_schema_fidelity}, only one feature-name mismatch was observed in the XGBoost setting with \textit{Claude-Sonnet-4.5}. In that case, the generated list included ``Number of Delinquencies in Past 2 Months'', instead of the valid feature ``Number of Delinquencies in Past 2 Years''. This error should be interpreted as a feature-name substitution rather than a broader formatting failure. We observed no duplicate features or malformed JSON responses. These findings suggest that violations of the constrained output schema were infrequent.

\begin{table}[htbp]
\centering
\caption{Schema-level fidelity check for RQ1 outputs}
\label{tab:rq1_schema_fidelity}
\begin{tabular}{lcccc}
\hline
\textbf{Base Model}
& \textbf{\shortstack{Total\\Outputs}}
& \textbf{\shortstack{Invalid \\ Features} }
& \textbf{\shortstack{Duplicate\\Features} }
& \textbf{\shortstack{Malformed\\JSON} }\\
\hline
Logistic Regression & 1500 & 0 & 0  & 0 \\
XGBoost & 1500 &1 & 0 & 0 \\
\hline
\end{tabular}
\end{table}

Conditional on schema-valid outputs, all RQ1 ranking-preservation metrics are equal to one across both baseline models and all evaluated LLMs. Specifically, Overlap@$K$ equals 1.000 for K = 5, 10, 15, and 20, indicating that the returned feature sets exactly match the supplied reference feature sets at each cutoff. Kendall's $\tau$ is also equal to 1.000 for all evaluated cutoffs, indicating that the relative ordering of the returned features exactly matches the supplied reference ranking.

Because all RQ1 ranking-preservation metrics are equal to one among schema-valid outputs, the empirical distributions are degenerate, with mean, quartiles, minimum, and maximum all equal to 1.000. The nonparametric bootstrap 95\% confidence intervals for the mean metrics are therefore also $[1.000, 1.000]$. These results suggest that, under the constrained translator setting, LLMs can reliably reproduce model-provided attribution rankings when the ranking is explicitly supplied in the prompt. In other words, the evaluated LLMs preserved both feature membership and feature order in all schema-valid RQ1 outputs, while only one schema-level feature-name substitution was observed across all 3000 runs in RQ1. Therefore, RQ1 provides evidence that LLMs are effective at reformatting existing model-grounded reference explanations.

\subsection{RQ2: LLM as Autonomous Explainer}

RQ2 evaluates whether LLMs can autonomously generate feature-importance rankings that align with model-grounded reference explanations when the reference attribution ranking is not provided in the prompt. Unlike RQ1, RQ2 examines a more challenging setting in which the LLM must infer explanatory importance from the instance-level input information and task instructions alone.

The evaluation is conducted under two experimental conditions. In the \textit{Top-4 condition}, the LLM is instructed to return only the four most important features. This setting is designed to approximate practical credit explanation scenarios, where explanations typically focus on a small number of primary reasons rather than a complete ordering of all input variables. In the \textit{full-ranking condition}, the LLM is instructed to return a complete ranking of all input features. This setting provides a stricter test of whether the LLM can recover the full model-grounded attribution structure.

For each condition, the LLM-generated ranking $\hat{\mathcal{T}}_i$ is compared with the reference ranking $\mathcal{T}_i$ derived from the corresponding baseline model. For the \textit{Top-4 condition}, we focus primarily on Overlap@4, as this metric directly reflects the practical credit explanation setting. For the \textit{full-ranking condition}, we evaluate both Overlap@K, with $K \in \{5,10,15,20\}$, and Kendall's $\tau$ to assess feature-set alignment and ordering consistency across longer rankings.

\subsubsection{Top-4 Condition}
\label{subsec:top4}
Tables~\ref{tab:rq2_top4_logistic} and~\ref{tab:rq2_top4_xgboost} show the Overlap@4 summary for the logistic regression and XGBoost models, respectively. We note that a small number of Gemini-2.5-Flash outputs were excluded because the responses were truncated before completing the required JSON object. This indicates that \textit{Gemini-2.5-Flash} was slightly less reliable in maintaining strict structured-output compliance, even though the requested Top-4 ranking format was relatively simple. To complement the summary statistics reported in the main text, the full distribution of Top-4 overlap counts under both zero-shot and few-shot prompting is provided in Appendix~\ref{sec:appendix_top4}. It reports the number of valid outputs with 0, 1, 2, 3, or 4 overlapping features for each baseline model and LLM setting.

In the logistic regression setting, all three LLMs show limited alignment with the coefficient-based reference explanations under zero-shot prompting, with mean Overlap@4 ranging from 0.064 to 0.199. Few-shot prompting improves Top-4 alignment for all three models, most notably for GPT-4-turbo and Claude-Sonnet-4.5. However, the improvement remains limited, as the mean Overlap@4 remains below 0.60 for all LLMs. This result indicates that few-shot prompting can partially recalibrate the LLMs' feature-selection behavior, but it does not substitute for direct access to model-based attributions.

For XGBoost, the zero-shot Overlap@4 values are higher than those observed for logistic regression. This pattern may be partly related to the stronger predictive performance of the XGBoost model. Since XGBoost achieves higher PR-AUC and KS than logistic regression, its SHAP-based rankings may reflect the attribution structure of a more discriminative fitted model, whereas logistic-regression contributions are constrained by a linear specification. Because zero-shot LLMs likely rely on general domain priors when selecting explanatory features, these priors may coincide more closely with XGBoost SHAP explanations than with logistic-regression contribution rankings. It suggests that the top SHAP features from the stronger model may be more consistent with general credit-risk heuristics.

The smaller improvement under the XGBoost few-shot condition further suggests that demonstration examples alone are insufficient for the LLMs to recover SHAP-based tree-model explanations. XGBoost attributions are highly instance-specific and may reflect nonlinear interactions among borrower characteristics. Therefore, two examples may not provide enough information for the LLMs to infer a stable attribution rule across borrowers. By contrast, logistic-regression contributions are more globally structured, which may explain why few-shot prompting produces a clearer improvement in the logistic regression setting.

Overall, the Top-4 results show that LLM-generated explanations achieve only partial alignment with model-grounded reference explanations. Although few-shot prompting improves overlap in several settings, the selected features often remain different from those identified by coefficient-based contributions or SHAP values. This finding suggests that LLMs should not be used as a substitute for quantitative attribution methods in adverse action analysis. Their outputs are not consistently grounded in the fitted model's instance-specific attribution structure and therefore introduce additional credit risk.

\begin{table}[htbp]
\centering
\caption{RQ2 Top-4 Condition: Overlap@4 for Logistic Regression}
\label{tab:rq2_top4_logistic}
\small
\setlength{\tabcolsep}{4pt}
\begin{tabular}{llccc}
\hline
\textbf{\shortstack{Prompt\\Setting}} 
& \textbf{\shortstack{LLM\\Model}} 
& \textbf{\shortstack{Sample\\($N$)}} 
& \textbf{\shortstack{Mean\\(95\% CI)}} 
& \textbf{\shortstack{Median\\(Min, Max)}} \\
\hline
\\[-10pt]
Zero-shot & GPT-4-turbo        & 500 & \shortstack{0.199\\(0.186, 0.212)} & \shortstack{0.25\\(0.00, 0.50)} \\
Zero-shot & Claude-Sonnet-4.5  & 500 & \shortstack{0.064\\(0.054, 0.075)} & \shortstack{0.00\\(0.00, 0.50)} \\
Zero-shot & Gemini-2.5-Flash   & 498 & \shortstack{0.067\\(0.057, 0.076)} & \shortstack{0.00\\(0.00, 0.25)} \\
\hline
\\[-10pt]
Few-shot  & GPT-4-turbo        & 500 & \shortstack{0.515\\(0.494, 0.529)} & \shortstack{0.50\\(0.00, 1.00)} \\
Few-shot  & Claude-Sonnet-4.5  & 500 & \shortstack{0.415\\(0.394, 0.436)} & \shortstack{0.50\\(0.00, 1.00)} \\
Few-shot  & Gemini-2.5-Flash   & 500 & \shortstack{0.182\\(0.166, 0.198)} & \shortstack{0.25\\(0.00, 0.75)} \\
\hline
\end{tabular}
\end{table}

\begin{table}[htbp]
\centering
\caption{RQ2 Top-4 Condition: Overlap@4 for XGBoost}
\label{tab:rq2_top4_xgboost}
\small
\setlength{\tabcolsep}{4pt}
\begin{tabular}{llccc}
\hline
\textbf{\shortstack{Prompt\\Setting}} 
& \textbf{\shortstack{LLM\\Model}} 
& \textbf{\shortstack{Sample\\($N$)}} 
& \textbf{\shortstack{Mean\\(95\% CI)}} 
& \textbf{\shortstack{Median\\(Min, Max)}} \\
\hline
\\[-10pt]
Zero-shot & GPT-4-turbo        & 500 & \shortstack{0.214\\(0.199, 0.248)} & \shortstack{0.25\\(0.00, 0.75)} \\
Zero-shot & Claude-Sonnet-4.5  & 500 & \shortstack{0.233\\(0.218, 0.248)} & \shortstack{0.25\\(0.00, 0.75)} \\
Zero-shot & Gemini-2.5-Flash   & 496 & \shortstack{0.102\\(0.090, 0.114)} & \shortstack{0.00\\(0.00, 0.25)} \\
\hline
\\[-10pt]
Few-shot  & GPT-4-turbo        & 500 & \shortstack{0.383\\(0.366, 0.400)} & \shortstack{0.25\\(0.00, 1.00)} \\
Few-shot  & Claude-Sonnet-4.5  & 500 & \shortstack{0.321\\(0.304, 0.339)} & \shortstack{0.25\\(0.00, 1.00)} \\
Few-shot  & Gemini-2.5-Flash   & 500 & \shortstack{0.204\\(0.188, 0.219)} & \shortstack{0.25\\(0.00, 0.75)} \\
\hline
\end{tabular}
\end{table}

\subsubsection{Full-ranking Condition}

As the full-ranking condition requires the LLM models to generate a complete 24-feature ranking under a relatively long constrained prompt, output-format compliance becomes a prerequisite for meaningful alignment evaluation. As shown by the prompt-length estimates in Appendix~\ref{appendix:token_length}, this setting imposes a substantially longer structured-generation requirement than RQ1. We therefore first evaluate schema-level validity by checking whether each response contains a complete ranking of all 24 input features, with no missing, invalid, or duplicated feature names, as shown in Table~\ref{tab:rq2_schema_fidelity}.

\textit{Gemini-2.5-Flash} shows substantially lower schema-level fidelity than the other two LLMs under the full-ranking condition. In many cases, Gemini returned only one or two ranked features rather than the required complete 24-feature list, despite using the same prompt template, parser, and evaluation pipeline as \textit{GPT-4-turbo} and \textit{Claude-Sonnet-4.5}. This pattern reflects the weaker adherence to long constrained-output instructions in this setting. Because \textit{Gemini-2.5-Flash} did not produce sufficient valid full-ranking outputs, the subsequent RQ2 full-ranking analysis focuses only on \textit{GPT-4-turbo} and \textit{Claude-Sonnet-4.5}.

\begin{table}[htbp]
\centering
\caption{Schema-level fidelity check for RQ2 Full-ranking}
\label{tab:rq2_schema_fidelity}
\begin{tabular}{lllcccc}
\hline
\textbf{\shortstack{Base \\ Model} }
& \textbf{\shortstack{Prompt\\ Setting} }
& \textbf{\shortstack{LLM \\Model} }
& \textbf{\shortstack{Valid \\Outputs}} 
& \textbf{\shortstack{Full\\ list} }
& \textbf{\shortstack{Invalid \\Features} }
& \textbf{\shortstack{Duplicate \\Features} }\\
\hline
\shortstack{Logistic \\Regression} & ZeroShot & GPT-4-turbo & 500 & 500 & 0 & 0 \\
\shortstack{Logistic \\Regression}  & ZeroShot & Claude-Sonnet-4.5 & 500 & 500 & 0 & 0 \\
\shortstack{Logistic \\Regression}  & ZeroShot & Gemini-2.5-Flash & 499 & 0 & 0 & 0 \\
\shortstack{Logistic \\Regression}  & FewShot & GPT-4-turbo & 500 & 469 & 0 & 0\\
\shortstack{Logistic \\Regression} & FewShot & Claude-Sonnet-4.5 & 500 & 500 & 0 & 0 \\
\shortstack{Logistic \\Regression}  & FewShot & Gemini-2.5-Flash & 500 & 0 & 0 & 0\\
\hline
XGBoost & ZeroShot & GPT-4-turbo & 500 & 500 & 0 & 0 \\
XGBoost& ZeroShot & Claude-Sonnet-4.5 & 500 & 500 & 0 & 0 \\
XGBoost & ZeroShot & Gemini-2.5-Flash & 498 & 0 & 0 & 0 \\
XGBoost & FewShot & GPT-4-turbo & 500 & 494 & 0 & 0\\
XGBoost & FewShot & Claude-Sonnet-4.5 & 500 & 500 & 0 & 0\\
XGBoost & FewShot & Gemini-2.5-Flash & 500 & 0 & 0 & 0\\
\hline
\end{tabular}
\end{table}

Tables~\ref{tab:rq2_topK_logistic} and~\ref{tab:rq2_topK_xgboost} report Overlap@K mean and bootstrap $95\%$ confidence interval with $K \in \{5,10,15,20\}$ under the \textit{full-ranking condition}. Because the full-ranking task requires each model to rank all 24 input features, larger $K$ values naturally produce higher overlap. Therefore, the smaller cutoffs, particularly O@5 and O@10, are more informative for evaluating whether the LLM recovers the most important model-grounded features.

For logistic regression, few-shot prompting substantially improves feature-set recovery. \textit{GPT-4-turbo} improves from 0.392 to 0.609 at mean O@5, while \textit{Claude-Sonnet-4.5} improves from 0.210 to 0.536. This suggests that demonstration examples help the LLMs better approximate the coefficient-based attribution structure, especially for the most highly ranked features. However, the overlap remains far from perfect, indicating that few-shot prompting only partially recovers the model-grounded ranking.

For XGBoost, the zero-shot O@5 values are close to the random-overlap baseline implied by the 24-feature ranking universe, suggesting limited recovery of the top SHAP-ranked features. Few-shot prompting improves performance, particularly for GPT-4-turbo, whose O@5 mean increases from 0.228 to 0.423. However, the improvement is smaller and less consistent than in the logistic regression setting. This pattern is consistent with our observations in Section \ref{subsec:top4}, where local attributions are more instance-specific and may reflect nonlinear feature interactions.

Overall, the full-ranking Overlap@K results indicate that LLMs can recover some high-importance features, especially under few-shot prompting, but they do not reliably reconstruct the complete model-grounded attribution structure.

\begin{table}[htbp]
\centering
\caption{RQ2 Full-Ranking Condition: Overlap@K for Logistic Regression}
\label{tab:rq2_topK_logistic}
\small
\setlength{\tabcolsep}{3.5pt}
\begin{tabular}{llccccc}
\hline
\textbf{Prompt} & \textbf{LLM} & \textbf{$N$}
& \textbf{O@5} & \textbf{O@10} & \textbf{O@15} & \textbf{O@20} \\
\hline
\\[-10pt]
Zero-shot & GPT-4-turbo & 500
& \shortstack{0.392\\(0.381, 0.403)}
& \shortstack{0.475\\(0.465, 0.486)}
& \shortstack{0.647\\(0.638, 0.655)}
& \shortstack{0.827\\(0.823, 0.832)} \\

Zero-shot & Claude-Sonnet-4.5 & 500
& \shortstack{0.210\\(0.197, 0.223)}
& \shortstack{0.537\\(0.528, 0.546)}
& \shortstack{0.773\\(0.766, 0.779)}
& \shortstack{0.892\\(0.889, 0.895)} \\
\hline
\\[-10pt]
Few-shot & GPT-4-turbo & 469
& \shortstack{0.609\\(0.594, 0.626)}
& \shortstack{0.670\\(0.663, 0.677)}
& \shortstack{0.769\\(0.764, 0.775)}
& \shortstack{0.947\\(0.944, 0.951)} \\

Few-shot & Claude-Sonnet-4.5 & 500
& \shortstack{0.536\\(0.519, 0.552)}
& \shortstack{0.665\\(0.658,0.673)}
& \shortstack{0.762\\(0.756,0.767)}
& \shortstack{0.904\\(0.900,0.907)} \\
\hline
\end{tabular}
\begin{flushleft}
\footnotesize
Note: O@K denotes Overlap@K. Values are reported as mean with bootstrap 95\% confidence intervals in parentheses.
\end{flushleft}
\end{table}

\begin{table}[htbp]
\centering
\caption{RQ2 Full-Ranking Condition: Overlap@K for XGBoost}
\label{tab:rq2_topK_xgboost}
\small
\setlength{\tabcolsep}{3.5pt}
\begin{tabular}{llccccc}
\hline
\textbf{Prompt} & \textbf{LLM} & \textbf{$N$}
& \textbf{O@5} & \textbf{O@10} & \textbf{O@15} & \textbf{O@20} \\
\hline
\\[-10pt]
Zero-shot & GPT-4-turbo & 500
& \shortstack{0.228\\(0.215, 0.241)}
& \shortstack{0.400\\(0.390, 0.410)}
& \shortstack{0.610\\(0.603, 0.616)}
& \shortstack{0.821\\(0.818, 0.824)} \\

Zero-shot & Claude-Sonnet-4.5 & 500
& \shortstack{0.242\\(0.230, 0.255)}
& \shortstack{0.487\\(0.478, 0.496)}
& \shortstack{0.667\\(0.661, 0.672)}
& \shortstack{0.859\\(0.857, 0.861)} \\
\hline
\\[-10pt]
Few-shot & GPT-4-turbo & 494
& \shortstack{0.423\\(0.409, 0.438)}
& \shortstack{0.536\\(0.527, 0.545)}
& \shortstack{0.686\\(0.680, 0.691)}
& \shortstack{0.892\\(0.889, 0.894)} \\

Few-shot & Claude-Sonnet-4.5 & 500
& \shortstack{0.315\\(0.298, 0.331)}
& \shortstack{0.543\\(0.534,0.551)}
& \shortstack{0.697\\(0.691,0.702)}
& \shortstack{0.887\\(0.885,0.890)} \\
\hline
\end{tabular}
\begin{flushleft}
\footnotesize
Note: O@K denotes Overlap@K. Values are reported as mean with bootstrap 95\% confidence intervals in parentheses.
\end{flushleft}
\end{table}

Table~\ref{tab:rq2_kendall_logistic} and \ref{tab:rq2_kendall_xgboost} show the corresponding Kendall's $\tau$ with different $K$ values. Overall, the Kendall's $\tau$ values remain modest across most settings, indicating that LLMs do not reliably recover the full attribution ordering even when selected high-important features overlap.

For logistic regression, few-shot prompting improves ordering consistency for both \textit{GPT-4-turbo} and \textit{Claude-Sonnet-4.5}. This pattern suggests that demonstration examples help LLMs approximate the more structured, additive attribution pattern of logistic regression. However, the resulting Kendall's $\tau$ values are under 0.4 and far from perfect, indicating that few-shot prompting only partially recovers the model-grounded ranking.

For XGBoost, the improvement from few-shot prompting is weaker and less consistent. \textit{GPT-4-turbo} shows some improvement at larger cutoffs, whereas \textit{Claude-Sonnet-4.5} exhibits little or no improvement under the few-shot setting. This finding is consistent with the nonlinear and interaction-driven nature of SHAP explanations for tree-based ensemble models, where local feature rankings may vary substantially across instances.

Taken together, the Kendall's $\tau$ results reinforce the conclusion from the Overlap@K analysis. LLMs may recover some relevant features, particularly under few-shot prompting, but they do not reliably reconstruct the complete model-grounded attribution structure. This distinction should be considered for credit risk governance because full-ranking consistency requires not only identifying plausible features but also preserving their relative importance according to the fitted predictive model.

\begin{table}[H]
\centering
\caption{RQ2 Full-Ranking Condition: Kendall's $\tau$ for Logistic Regression}
\label{tab:rq2_kendall_logistic}
\small
\setlength{\tabcolsep}{3.5pt}
\begin{tabular}{llccccc}
\hline
\textbf{Prompt} & \textbf{LLM} & \textbf{$N$}
& \textbf{$\tau$@5} & \textbf{$\tau$@10} & \textbf{$\tau$@15} & \textbf{$\tau$@20} \\
\hline
\\[-10pt]
Zero-shot & GPT-4-turbo & 500
& \shortstack{0.060\\(0.018, 0.104)}
& \shortstack{0.171\\(0.140, 0.201)}
& \shortstack{0.083\\(0.071, 0.095)}
& \shortstack{0.126\\(0.115, 0.136)} \\

Zero-shot & Claude-Sonnet-4.5 & 500
& \shortstack{-0.108\\(-0.137, -0.078)}
& \shortstack{-0.238\\(-0.260, -0.216)}
& \shortstack{-0.113\\(-0.125, -0.102)}
& \shortstack{0.159\\(0.150, 0.168)} \\
\hline
\\[-10pt]
Few-shot & GPT-4-turbo & 469
& \shortstack{0.096\\(0.051, 0.142)}
& \shortstack{0.287\\(0.266, 0.308)}
& \shortstack{0.365\\(0.352, 0.378)}
& \shortstack{0.399\\(0.388, 0.409)} \\

Few-shot & Claude-Sonnet-4.5 & 500
& \shortstack{-0.144\\(-0.185, -0.101)}
& \shortstack{0.123\\(0.100, 0.147)}
& \shortstack{0.260\\(0.247, 0.274)}
& \shortstack{0.358\\(0.348, 0.366)} \\
\hline
\end{tabular}
\begin{flushleft}
\footnotesize
Note: $\tau$@K denotes Kendall's $\tau$ computed over the intersection of the two Top-K ranked features. Values are reported as mean with bootstrap 95\% confidence intervals in parentheses.
\end{flushleft}
\end{table}

\begin{table}[htbp]
\centering
\caption{RQ2 Full-Ranking Condition: Kendall's $\tau$ for XGBoost}
\label{tab:rq2_kendall_xgboost}
\small
\setlength{\tabcolsep}{3.5pt}
\begin{tabular}{llccccc}
\hline
\textbf{Prompt} & \textbf{LLM} & \textbf{$N$}
& \textbf{$\tau$@5} & \textbf{$\tau$@10} & \textbf{$\tau$@15} & \textbf{$\tau$@20} \\
\hline
\\[-10pt]
Zero-shot & GPT-4-turbo & 500
& \shortstack{0.247\\(0.215, 0.279)}
& \shortstack{0.160\\(0.137, 0.183)}
& \shortstack{0.056\\(0.039, 0.071)}
& \shortstack{0.068\\(0.054, 0.081)} \\

Zero-shot & Claude-Sonnet-4.5 & 500
& \shortstack{0.217\\(0.182, 0.251)}
& \shortstack{0.256\\(0.238, 0.274)}
& \shortstack{0.166\\(0.153, 0.179)}
& \shortstack{0.115\\(0.103, 0.126)} \\
\hline
\\[-10pt]
Few-shot & GPT-4-turbo & 494
& \shortstack{0.204\\(0.170, 0.237)}
& \shortstack{0.247\\(0.228, 0.266)}
& \shortstack{0.210\\(0.196, 0.224)}
& \shortstack{0.154\\(0.144, 0.164)} \\

Few-shot & Claude-Sonnet-4.5 & 500
& \shortstack{0.262\\(0.227, 0.298)}
& \shortstack{0.176\\(0.153, 0.199)}
& \shortstack{0.150\\(0.134, 0.166)}
& \shortstack{0.115\\(0.101, 0.129)} \\
\hline
\end{tabular}
\begin{flushleft}
\footnotesize
Note: $\tau$@K denotes Kendall's $\tau$ computed over the intersection of the two Top-K ranked features. Values are reported as mean with bootstrap 95\% confidence intervals in parentheses.
\end{flushleft}
\end{table}

\section{Conclusions}
\label{SecConclusion}

This study evaluates whether large language models (LLMs) can serve as post-hoc explainability tools for credit risk models, addressing two research questions: whether LLMs can preserve supplied feature-importance rankings under strict output constraints when reference attributions are provided (RQ1), and whether they can autonomously generate aligned feature-importance rankings without access to such references (RQ2). Using logistic regression and XGBoost as baseline classifiers, we compare LLM-generated rankings with coefficient-based and SHAP-based reference explanations using Top-K feature overlap and Kendall's $\tau$.

The results show a clear distinction between the translator and autonomous-explainer settings. Under RQ1, where the reference attribution ranking is explicitly supplied in the prompt, the evaluated LLMs exhibit very strong structural fidelity. Across 3,000 outputs, only one schema-level feature-name substitution was observed, and no duplicate features or malformed JSON responses were found. Conditional on schema-valid outputs, all ranking-preservation metrics are equal to one across all baseline models, LLMs, and evaluated cutoffs. These findings suggest that, under a highly constrained prompt design, LLMs can reliably reproduce supplied model-grounded attribution rankings. However, this result should be interpreted as evidence of ranking-preservation and structured-output compliance, rather than evidence that the LLM independently understands or verifies the underlying model reasoning.

The RQ2 results provide a more cautious picture. When reference attributions are not supplied, LLM-generated rankings show only partial alignment with model-grounded reference explanations. In the Top-4 condition, few-shot prompting improves alignment in several settings, especially for logistic regression, but the mean Overlap@4 remains below a level that would support using LLMs as standalone explanation tools. In the XGBoost setting, zero-shot and few-shot outputs also show limited agreement with SHAP-based reference explanations, suggesting that LLMs have difficulty recovering instance-specific, nonlinear, and interaction-driven attribution structures from borrower information alone.

The full-ranking condition further reinforces this conclusion. Although \textit{GPT-4-turbo} and \textit{Claude-Sonnet-4.5} can generate complete 24-feature rankings with relatively strong schema compliance, \textit{Gemini-2.5-Flash} frequently fails to produce a valid full ranking under the longer constrained-output setting. Among valid full-ranking outputs, few-shot prompting improves Overlap@K, particularly for logistic regression and for \textit{GPT-4-turbo}, but the recovered rankings remain far from perfect. Kendall's $\tau$ values remain modest across most settings, indicating that even when LLMs identify some overlapping high-importance features, they do not reliably preserve the relative ordering of the full model-grounded attribution structure.

These findings have direct implications for credit risk model governance. LLMs should not be viewed as substitutes for formal post-hoc attribution methods such as coefficient-based decomposition or SHAP. Model-based explainers derive attributions from the fitted model structure, whereas LLMs infer feature importance from prompt-level information, learned language patterns, and general domain priors. As a result, autonomous LLM explanations may select plausible credit-risk factors without faithfully reflecting the actual prediction algorithm of the underlying model. This distinction is particularly important in high-stakes credit settings, where explanations must be traceable, auditable, and consistent with the approved model.

Several limitations motivate future research. First, the evaluation is conducted on a relatively small sample due to computational cost constraints and focuses on rank alignment rather than user-centered outcomes. Larger-scale empirical analyses are therefore needed in future research to examine whether the observed findings remain stable across broader samples, additional credit datasets, and more diverse model and prompting configurations.
Second, the current design does not include perturbation-based stress tests, such as deliberately shuffled attribution rankings or decoy features. Such tests would be most informative in a setting where the LLM receives both raw feature values and a supplied attribution ranking, because they could reveal whether the model follows the provided ranking or overrides it based on its own reasoning from feature values. Third, while alignment with post-hoc attributions is informative, such alignment does not by itself guarantee that explanations are faithful, actionable, or legally sufficient, particularly for complex models such as XGBoost. Future work should evaluate whether the observed LLM alignment patterns remain stable when using stronger production-grade credit models, additional datasets, and calibrated probability outputs designed for operational decision.

Overall, our findings support a constrained and governance-oriented use of LLMs in credit explainability workflows. LLMs can reliably restate supplied attribution rankings under strict output constraints, but they do not consistently reconstruct model-grounded reference explanations when operating autonomously. In practical model risk management workflows, LLMs may be useful as narrative or formatting interfaces that translate quantitative model-grounded attributions into clearer, more accessible explanations for analysts, validators, or other stakeholders.

\appendix
\section{Prompt Templates}
\label{sec:prompt}
\subsection{RQ1: LLM as Translator Prompt}
\label{subsec:RQ1_prompt}

The prompt template below was used for all three LLMs evaluated in the RQ1 experiment, including \textit{GPT-4-turbo}, \textit{Claude-Sonnet-4.5}, and \textit{Gemini-2.5-Flash}. The template shown corresponds to the logistic regression model. For the XGBoost model, the same template was used, except that the model name ``Logistic Regression'' was replaced with ``XGBoost''. The input structure, \texttt{attribution\_table}, and the required JSON output schema, \texttt{OUTPUT\_SCHEMA}, are described in Appendix~\ref{subsec:Input_and_output}.

\subsubsection*{System Message}

\begin{lstlisting}[style=promptstyle]
You are given a Logistic Regression model-provided feature importance list.

The features are already ordered from most important to least important.

Your task is to reproduce the feature order exactly as provided.
Rules:
1. Copy the feature names in the exact order shown.
2. Do not reorder features based on feature meaning or credit-risk knowledge.
3. Do not add, remove, rename, merge, or explain features.
4. Return ONLY valid JSON, with no Markdown fences, prose, comments, or explanation.
5. Use exactly this schema:
{json.dumps(OUTPUT_SCHEMA, indent=2)}
\end{lstlisting}

\subsubsection*{User Context}
\begin{lstlisting}[style=promptstyle]
JSON OUTPUT ONLY

MODEL-PROVIDED FEATURE IMPORTANCE LIST:
The features below are already ordered from most important to least important.

{sample['attribution_table']}

Task:
Return the feature names in exactly the same order as shown above.
\end{lstlisting}

\subsection{RQ2: LLM as Autonomous Explainer Zero-shot Prompt}
\label{subsec:RQ2_prompt}

The prompt template below was used for all three LLMs evaluated in the RQ2 zero-shot experiment. The input structure and required JSON output schema are described in Appendix~\ref{subsec:Input_and_output}. The global variable \texttt{TOP\_N\_FEATURES} is included in both the system and user message to control the number of features returned in the output.

\subsubsection*{System Message}
\begin{lstlisting}[style=promptstyle]
You are an expert credit risk analyst.

Your task is to infer which borrower input features most likely contributed to the Logistic Regression model prediction.
Rules:
1. Use ONLY the borrower input values provided below.
2. Do not use attribution values, SHAP values, or a model-provided ranking; none are shown.
3. Return the top {TOP_N_FEATURES} features from most important to least important.
4. Copy feature names exactly as listed in the provided feature list.
5. Do not invent, rename, merge, or explain features.
6. Return ONLY valid JSON, with no Markdown fences, prose, comments, or explanation.
7. Use exactly this schema:
{json.dumps(OUTPUT_SCHEMA, indent=2)}
\end{lstlisting}

\subsubsection*{User Context}
\begin{lstlisting}[style=promptstyle]
PREDICTED DEFAULT PROBABILITY: {sample['predicted_probability']:.4f}

BORROWER INPUT VALUES:
{sample['input_table']}

Task:
Rank the top {TOP_N_FEATURES} features that most likely contributed to this Logistic Regression model prediction.
\end{lstlisting}

\subsection{RQ2: LLM as Autonomous Explainer Few-shot Prompt}

The prompt template below was used for all three LLMs evaluated in the RQ2 few-shot experiment. Compared with zero-shot prompt in Appendix~\ref{subsec:RQ2_prompt}, the only difference is the inclusion of two demonstration examples, which are described in Appendix~\ref{subsec:few_shot_example}. The input structure and required JSON output schema are described in Appendix~\ref{subsec:Input_and_output}. The global variable \texttt{TOP\_N\_FEATURES} is included in both the system and user message to control the number of features returned in the output.

\subsubsection*{System Message}
\begin{lstlisting}[style=promptstyle]
You are an expert credit risk analyst.

You will see two labeled examples: one training example with the highest model-predicted default probability and one with the lowest model-predicted default probability.
Each example shows borrower input values and the correct top Logistic Regression model features.
Then you will rank the top {TOP_N_FEATURES} features for a new borrower.
Rules:
1. Use ONLY the target borrower's input values.
2. Do not use attribution values, SHAP values, or a model-provided ranking for the target; none are shown.
3. Return the top {TOP_N_FEATURES} features from most important to least important.
4. Choose feature names exactly as listed in the provided feature list.
5. Do not invent, rename, merge, or explain features.
6. Return ONLY valid JSON, with no Markdown fences, prose, comments, or explanation.
7. Use exactly this schema:
{json.dumps(OUTPUT_SCHEMA, indent=2)}
\end{lstlisting}

\subsubsection*{User Context}
\begin{lstlisting}[style=promptstyle]
FEW-SHOT EXAMPLES:
{example_text}

TARGET BORROWER PREDICTED DEFAULT PROBABILITY: {sample['predicted_probability']:.4f}

TARGET BORROWER INPUT VALUES:
{sample['input_table']}

Task:
Rank the top {TOP_N_FEATURES} features that most likely contributed to this Logistic Regression model prediction.
\end{lstlisting}

\subsection{Input and Output Structure}
\label{subsec:Input_and_output}

\subsubsection*{Input Structure for RQ1}
\label{subsubsec: Input_RQ1}

The input \texttt{attribution\_table} for one representative RQ1 example is shown below.
\begin{lstlisting}[style=promptstyle]
Highest FICO Score
Lowest FICO Score
Loan Sub Grade
Annual Income
Monthly Payment
Loan Amount
Revolving Balance
Borrower Area
Total Number of Accounts
Borrower State
Debt to Income Ratio
Loan Grade
Employment Length
Income Verification Status
Number of Open Accounts
Revolving Utilization Rate
Interest Rate
Home Ownership Status
Loan Purpose
Number of Payments
Number of Delinquencies in Past 2 Years
Number of Public Bankruptcies
Number of Inquiries in Last 6 Months
Number of Derogatory Public Records
\end{lstlisting}

\subsubsection*{Input Structure for RQ2}
\label{subsubsec: Input_RQ2}

The input \texttt{attribution\_table} for one representative RQ2 example is shown below. The same input structure was used in both the zero-shot and few-shot settings.
\begin{lstlisting}[style=promptstyle]
Feature                                  | Input Value
---------------------------------------------------------
Home Ownership Status                    | RENT
Income Verification Status               | Not Verified
Loan Purpose                             | vacation
Borrower Area                            | West
Borrower State                           | CA
Number of Payments                       | 36
Loan Grade                               | B
Loan Sub Grade                           | B3
Employment Length                        | 2 years
Number of Public Bankruptcies            | 0
Loan Amount                              | 2000
Interest Rate                            | 11.7%
Monthly Payment                          | 66.16
Annual Income                            | 30,000
Debt to Income Ratio                     | 16.12
Number of Delinquencies in Past 2 Years  | 0
Lowest FICO Score                        | 680
Highest FICO Score                       | 684
Number of Inquiries in Last 6 Months     | 0
Number of Open Accounts                  | 5
Number of Derogatory Public Records      | 0
Revolving Balance                        | 18608
Revolving Utilization Rate               | 75.3%
Total Number of Accounts                 | 8
\end{lstlisting}

\subsubsection*{Output Structure}
\label{subsubsec: Output}
The following JSON output structure was imposed as a strict output constraint across all experimental settings.

\begin{lstlisting}[style=promptstyle]
OUTPUT_SCHEMA = {
    "ranked_features": [
        "feature_name_1",
        "feature_name_2",
        "feature_name_3",
    ]
}
\end{lstlisting}

\subsection{Few Shot Examples}
\label{subsec:few_shot_example}

\subsubsection*{Few-shot examples for Logistic Regression}
The following examples are the few-shot demonstrations included in the RQ2 prompt for the logistic regression model. We selected two extreme cases, corresponding to observations with the highest and lowest predicted probabilities of default, because they provide clear and contrasting attribution patterns. These examples are intended to familiarize the LLM with the input structure and to illustrate how feature-attribution information should be interpreted and translated into the required JSON response.

\begin{lstlisting}[style=promptstyle,
]
EXAMPLE 1 (Highest predicted probability training example):

BORROWER INPUT VALUES:
Feature                                  | Input Value
----------------------------------------------------------
Home Ownership Status                    | MORTGAGE
Income Verification Status               | Verified
Loan Purpose                             | home_improvement
Borrower Area                            | Northeast
Borrower State                           | NJ
Number of Payments                       | 60
Loan Grade                               | F
Loan Sub Grade                           | F5
Employment Length                        | 10+ years
Number of Public Bankruptcies            | 0
Loan Amount                              | 25000
Interest Rate                            | 21.7%
Monthly Payment                          | 686.8
Annual Income                            | 82,400
Debt to Income Ratio                     | 6.9
Number of Delinquencies in Past 2 Years  | 0
Lowest FICO Score                        | 745
Highest FICO Score                       | 749
Number of Inquiries in Last 6 Months     | 8
Number of Open Accounts                  | 3
Number of Derogatory Public Records      | 0
Revolving Balance                        | 0
Revolving Utilization Rate               | 0.0%
Total Number of Accounts                 | 17

CORRECT TOP FEATURES:
1. Loan Amount
2. Monthly Payment
3. Loan Sub Grade
4. Number of Inquiries in Last 6 Months

EXAMPLE 2 (Lowest predicted probability training example):

BORROWER INPUT VALUES:
Feature                                  | Input Value
----------------------------------------------------------
Home Ownership Status                    | MORTGAGE
Income Verification Status               | Source Verified
Loan Purpose                             | home_improvement
Borrower Area                            | West
Borrower State                           | CA
Number of Payments                       | 36
Loan Grade                               | C
Loan Sub Grade                           | C1
Employment Length                        | 10+ years
Number of Public Bankruptcies            | 0
Loan Amount                              | 5000
Interest Rate                            | 12.7%
Monthly Payment                          | 167.8
Annual Income                            | 6,000,000
Debt to Income Ratio                     | 0.01
Number of Delinquencies in Past 2 Years  | 0
Lowest FICO Score                        | 710
Highest FICO Score                       | 714
Number of Inquiries in Last 6 Months     | 1
Number of Open Accounts                  | 2
Number of Derogatory Public Records      | 0
Revolving Balance                        | 1434
Revolving Utilization Rate               | 37.7%
Total Number of Accounts                 | 10

CORRECT TOP FEATURES:
1. Annual Income
2. Highest FICO Score
3. Lowest FICO Score
4. Loan Sub Grade

\end{lstlisting}

\subsubsection*{Few-shot examples for XGBoost}
The following examples are the few-shot demonstrations included in the RQ2 prompt for the XGBoost model. Similarly, we selected two extreme cases, corresponding to observations with the highest and lowest predicted probabilities of default, to familiarize the LLM with the input structure and to illustrate how XGBoost-based feature-attribution information should be interpreted.

\begin{lstlisting}[style=promptstyle,
]
EXAMPLE 1 (Highest predicted probability training example):

BORROWER INPUT VALUES:
Feature                                  | Input Value
--------------------------------------------------------------
Home Ownership Status                    | OWN
Income Verification Status               | Source Verified
Loan Purpose                             | small_business
Borrower Area                            | Midwest
Borrower State                           | MI
Number of Payments                       | 60
Loan Grade                               | C
Loan Sub Grade                           | C1
Employment Length                        | Unknown
Number of Public Bankruptcies            | 0
Loan Amount                              | 16000
Interest Rate                            | 12.2%
Monthly Payment                          | 357.8
Annual Income                            | 40,000
Debt to Income Ratio                     | 8.88
Number of Delinquencies in Past 2 Years  | 0
Lowest FICO Score                        | 765
Highest FICO Score                       | 769
Number of Inquiries in Last 6 Months     | 7
Number of Open Accounts                  | 6
Number of Derogatory Public Records      | 0
Revolving Balance                        | 102318
Revolving Utilization Rate               | 15.3%
Total Number of Accounts                 | 13

CORRECT TOP FEATURES:
1. Loan Purpose
2. Revolving Balance
3. Annual Income
4. Number of Inquiries in Last 6 Months

EXAMPLE 2 (Lowest predicted probability training example):

BORROWER INPUT VALUES:
Feature                                  | Input Value
--------------------------------------------------------------
Home Ownership Status                    | MORTGAGE
Income Verification Status               | Not Verified
Loan Purpose                             | major_purchase
Borrower Area                            | Midwest
Borrower State                           | MI
Number of Payments                       | 36
Loan Grade                               | A
Loan Sub Grade                           | A1
Employment Length                        | 10+ years
Number of Public Bankruptcies            | 0
Loan Amount                              | 3500
Interest Rate                            | 5.4%
Monthly Payment                          | 105.6
Annual Income                            | 112,000
Debt to Income Ratio                     | 7.69
Number of Delinquencies in Past 2 Years  | 0
Lowest FICO Score                        | 800
Highest FICO Score                       | 804
Number of Inquiries in Last 6 Months     | 0
Number of Open Accounts                  | 7
Number of Derogatory Public Records      | 0
Revolving Balance                        | 2327
Revolving Utilization Rate               | 4.7%
Total Number of Accounts                 | 23

CORRECT TOP FEATURES:
1. Interest Rate
2. Annual Income
3. Revolving Utilization Rate
4. Loan Purpose

\end{lstlisting}










\section{Complementary Tables for RQ2 Top-4 Evaluation}
\label{sec:appendix_top4}


Tables~\ref{tab:rq2_zeroshot_count} and~\ref{tab:rq2_fewshot_count} report the distribution of Top-4 overlap counts under zero-shot and few-shot prompting for RQ2 Top-4 condition, respectively. Specifically, each table shows the number of valid outputs in which the LLM-generated Top-4 feature list shares 0, 1, 2, 3, or 4 features with the model-grounded reference Top-4 list. These tables complement the main results by showing whether the reported overlap mean reflects broad moderate alignment across instances or is driven by a smaller number of highly aligned cases.

\begin{table}[htbp]
\centering
\caption{Distribution of Feature-Overlap Counts in the RQ2 Top-4 Condition with Zero-Shot Setting}
\label{tab:rq2_zeroshot_count}
\begin{tabular}{llcccccc}
\hline
\textbf{Base Model} & \textbf{LLM Model} & \textbf{N} & \textbf{0 hits} & \textbf{1 hits} & \textbf{2 hits} & \textbf{3 hits} & \textbf{4 hits} \\
\hline
Logistic Regression & GPT-4-turbo & 500 & 155 & 293 & 52 & 0 & 0 \\
Logistic Regression & Claude-Sonnet-4.5 & 500 & 381 & 110 & 9 & 0 & 0 \\
Logistic Regression & Gemini-2.5-Flash & 498 & 366 & 132 & 0 & 0 & 0 \\
\hline
XGBoost & GPT-4-turbo & 500 & 151 & 275 & 69 & 5 & 0 \\ 
XGBoost & Claude-Sonnet-4.5 & 500 & 133 & 273 & 89 & 5 & 0 \\
XGBoost & Gemini-2.5-Flash & 496 & 303 & 184 & 9 & 0 & 0 \\
\hline
\end{tabular}
\end{table}

\begin{table}[H]
\centering
\caption{Distribution of Feature-Overlap Counts in the RQ2 Top-4 Condition with Few-Shot Setting}
\label{tab:rq2_fewshot_count}
\begin{tabular}{llcccccc}
\hline
\textbf{Base Model} & \textbf{LLM Model} & \textbf{N} & \textbf{0 hits} & \textbf{1 hits} & \textbf{2 hits} & \textbf{3 hits} & \textbf{4 hits} \\
\hline
Logistic Regression & GPT-4-turbo & 500 & 13 & 92 & 267 & 115 & 13 \\
Logistic Regression & Claude-Sonnet-4.5 & 500 & 79 & 104 & 229 & 85 & 3 \\
Logistic Regression & Gemini-2.5-Flash & 500 & 211 & 217 & 70 & 2 & 0 \\
\hline
XGBoost & GPT-4-turbo & 500 & 38 & 220 & 183 & 56 & 3 \\    
XGBoost & Claude-Sonnet-4.5 & 500 & 79 & 237 & 149 & 33 & 2 \\
XGBoost & Gemini-2.5-Flash & 500 & 168 & 262 & 65 & 5 & 0 \\
\hline
\end{tabular}
\end{table}

\section{Token Length}
\label{appendix:token_length}
For each experimental design, we estimate the mean of prompt length among 500 samples in Table \ref{tab:toke_length} using the \texttt{cl100k\_base} tokenizer as a consistent cross-model approximation. This tokenizer provides a standardized way to approximate how input text is segmented into tokens, allowing prompt lengths to be compared consistently across experimental settings. For RQ2, although both the \textit{Top-4 condition} and the \textit{full-ranking condition} are evaluated, their prompts differ only in the requested ranking features number. Hence, the prompt length is the same.

\begin{table}[htbp]
\centering
\caption{Token Length Estimate}
\label{tab:toke_length}
\begin{tabular}{lllccc}
\hline
\textbf{Base Model} & \textbf{Experiment} &   \textbf{Condition} & \textbf{System Mean} 
& \textbf{User Mean} & \textbf{Total Mean}\\
\hline
Logistic Regression & RQ1 & N/A & 136 & 161 & 297  \\
Logistic Regression & RQ2 & Zero Shot & 160 & 278 & 438  \\
Logistic Regression & RQ2 & Few Shot & 204 & 873 & 1077  \\
\hline
XGBoost  & RQ1 & N/A & 137 & 161 & 298  \\
XGBoost& RQ2 & Zero Shot & 161 & 279 & 440  \\
XGBoost & RQ2 & Few Shot & 205 & 862 & 1067 \\
\hline
\end{tabular}
\end{table}

\section*{Declaration of generative AI and AI-assisted technologies in the manuscript preparation process}
During the preparation of this work, the authors used \textit{ChatGPT 5.2} for manuscript rephrasing only. \textit{Codex 5.5 medium} was also used to help organize and polish the project code for public release, including improving code readability, comments, and repository structure. No AI tools were used to design the experiments, execute the statistical analyses, produce the reported results, or substantively interpret the findings. After using these AI tools, the authors reviewed and edited the content and code as needed.

\section*{Conflict of Interest Disclosure}
The authors declare that they have no conflict of interest regarding the 
publication of this paper. The views expressed in this article are those of the authors and do not necessarily represent the views of Citigroup Inc. or its affiliates.

\bibliographystyle{plainnat}
\bibliography{refs}
\end{document}